\documentclass[lettersize,journal]{IEEEtran}
\usepackage{amssymb}
\usepackage{amsmath,amsfonts}
\usepackage{algorithmic}
\usepackage{algorithm}
\usepackage{array}
\usepackage[caption=false,font=normalsize,labelfont=sf,textfont=sf]{subfig}
\usepackage{textcomp}
\usepackage{stfloats}
\usepackage{url}
\usepackage{verbatim}
\usepackage{graphicx}
\usepackage{cite}
\usepackage{epstopdf}

\newenvironment{proof}[1][Proof]{\indent\emph{#1:} }{\hfill\ \rule{0.5em}{0.5em}}

\newtheorem{lemma}{Lemma}
\newtheorem{proposition}{Proposition}
\newtheorem{corollary}{Corollary}
\newtheorem{fact}{Fact}
\newtheorem{definition}{Definition}
\newtheorem{remark}{Remark}

\newtheorem{assumption}{Assumption}

\def\begcen{\begin{center}}
\def\endcen{\end{center}}

\newcommand{\col}{ \mbox{col} }

\def\caly{{\cal Y}}
\def\calg{{\cal G}}

\def\calj{{\cal J}}

\def\calf{{\cal F}}

\def\hal{{1 \over 2}}

\def\liminf{\lim_{t \to \infty}}

\def\L2{{\cal L}_2}
\def\L2e{{\cal L}_{2e}}

\def\rea{\mathbb{R}}

\def\adj{\mbox{adj}}

\def\diag{\mbox{diag}}

\def\begmat#1{\begin{bmatrix}#1\end{bmatrix}}
\def\begali#1{\begin{align}{#1}\end{align}}
\def\begalis#1{\begin{align*}{#1}\end{align*}}

\def\begequarr{\begin{eqnarray}}
\def\endequarr{\end{eqnarray}}
\def\begequarrs{\begin{eqnarray*}}
\def\endequarrs{\end{eqnarray*}}
\def\begarr{\begin{array}}
\def\endarr{\end{array}}
\def\begequ{\begin{equation}}
\def\endequ{\end{equation}}
\def\lab{\label}
\def\begdes{\begin{description}}
\def\enddes{\end{description}}
\def\begenu{\begin{enumerate}}
\def\begite{\begin{itemize}}
\def\endite{\end{itemize}}
\def\endenu{\end{enumerate}}

\def\lef[{\left[\begin{array}}
\def\rig]{\end{array}\right]}

\def\begcen{\begin{center}}
\def\endcen{\end{center}}
\def\begrem{\begin{remark}\rm}
\def\endrem{\end{remark}}
\def\begassum{\begin{assumption}}
\def\endassum{\end{assumption}}
\def\begassums{\begin{assumption*}}
\def\endassums{\end{assumption*}}
\def\begassu{\begin{ass}}
\def\endassu{\end{ass}}
\def\beglem{\begin{lemma}}
\def\endlem{\end{lemma}}
\def\begcor{\begin{corollary}}
\def\endcor{\end{corollary}}
\def\begfac{\begin{fact}}
\def\endfac{\end{fact}}

\def\liminf{\lim_{t \to \infty}}

\def\L2e{{\cal L}_{2e}}

\def\rea{\mathbb{R}}

\def\diag{\mbox{diag}}

\def\adj{\mbox{adj}}
\def\col{\mbox{col}}
\def\hal{{1 \over 2}}

\def\diag{\mbox{diag}}

\def\IJC{{\it Int. J. of Control}}

\def\AUT{{\it Automatica}}

\def\begsubequ{\begin{subequations}}
	\def\endsubequ{\end{subequations}}
\def\begpro{\begin{proposition}}
	\def\endpro{\end{proposition}}
\def\beglem{\begin{lemma}}
	\def\endlem{\end{lemma}}
\def\begass{\begin{assumption}}
	\def\endass{\end{assumption}}
\def\begcor{\begin{corollary}}
	\def\endcor{\end{corollary}}
\def\begproo{\begin{proof}}
	\def\endproo{\end{proof}}
\newcommand{\idot}{\stackrel{\dot {\frown}} {i}}
\usepackage{xcolor}

\begin{document}

\title{Interturn Fault Detection in IPMSMs:\\ Two Adaptive Observer-based Solutions}
\author{Romeo Ortega, \emph{Life Fellow}, IEEE, Alexey Bobtsov, \emph{Senior Member}, IEEE,\\ Leyan Fang, Oscar Texis-Loaiza, and Johannes Schiffer \thanks{%
\emph{(Corresponding author: Leyan Fang.)}%
} 
\thanks{%
Romeo Ortega is with the Departamento Acad\'{e}mico de Sistemas Digitales,
ITAM, Mexico City 01080, Mexico (email: romeo.ortega@itam.mx).}
\thanks{%
Alexey Bobtsov is with the Faculty of Control System and Robotics, ITMO University, 197101 Saint Petersburg, Russia (email: bobtsov@mail.ru).} 
\thanks{%
Leyan Fang is with the Center for Control Theory and Guidance Technology, Harbin Institute of Technology, Harbin
150001, China and also with the Departamento Acad\'{e}mico de Sistemas Digitales,
ITAM, Mexico City 01080, Mexico (email: leyan.fang@itam.mx).}
\thanks{%
Oscar Texis-Loaiza and Johannes Schiffer are with the Control Systems and Network Control Technology Group, Brandenburg University of Technology CottbusSenftenberg (BTU C-S), 03046 Cottbus, Germany (e-mail: texisosc@b-tu.de; e-mail: Schiffer
@b-tu.de).} }

\maketitle

\begin{abstract}
In this paper we address the problem of online detection of inter-turn short-circuit faults (ITSCFs) that occur in permanent magnet synchronous motors (PMSMs). {We} propose two solutions to this problem: (i) a very simple linear observer and (ii) a generalized parameter estimation based observer, that incorporates a high performance estimator---with both observers detecting the short-circuit current and the fault intensity. Although the first solution guarantees the detection of the fault exponentially fast, the rate of convergence is fully determined by the motor parameters that, in some cases, may be too slow. The second observer, on the other hand, ensures {\em finite convergence time} under the weakest assumption of interval excitation. To make the observers {\em adaptive}, we develop a parameter estimator that, in the case of isotropic  PMSMs, estimates on-line (exponentially fast) the resistance and inductance of the motor. It should be underscored that, in contrast with existing observers (including the widely popular Kalman filter) that provide {\em indirect} information of the fault current, our observers provide {\em explicit} one---namely the amplitude of the fault current. The performance of both observers, in their linear and generalized parameter estimation-based versions, is illustrated with realistic simulation studies. 
\end{abstract}


\section{Introduction}

\IEEEPARstart{P}{ermanent} magnet synchronous motors (PMSMs) largely dominate various application areas, particularly in the realm of electric vehicles \cite{atallah2011,lazari2014,CHOetal,NAMbook}. Despite their remarkable advantages, PMSMs are highly susceptible to electrical faults. Among the different types of faults affecting PMSMs (including stator winding faults, rotor faults, and power electronics faults), inter-turn short-circuit faults (ITSCFs) are the most commonly encountered ones \cite{DUetal,ZHAetal}. When an ITSCF occurs, a high circulating current is generated in the shorted circuit \cite{ZAFetal,BOUetal}. If the ITSCF is not diagnosed in a timely manner, it may propagate and lead to more severe faults, such as phase-to-phase faults, phase-to-ground faults, or demagnetization faults, for further details on the ITSCF detection problem see \cite{CHEetal}.
	
Fault diagnosis methods are generally classified into two categories: on-line and off-line approaches. On-line methods have gained considerable attention because they offer real-time feedback. In the context of PMSM fault detection, on-line diagnostic techniques are typically divided into three main groups: signal processing-based techniques, artificial intelligence-based approaches, and analytical model-based methods \cite{zhang2022fault}. The use of the qualifier ``model-based" has to be placed in its true context. Indeed, although a rather precise mathematical description of the appearance of the ITSCF, to the best of the authors' knowledge, this model has been used in one of the two following ways.\\

\noindent[{\bf W1}] The derivation of a two dimensional relation between the signals of interest of the form ${d i_{dq}(t) \over dt}=a(t)+b(t)$, where we are interested in estimating $i_{dq}(t) \in \rea^2$ and we assume $a(t)$ {\em known} and $b(t)$ {\em unknown}. High-gain injection methods are then used to reconstruct  $i_{dq}(t)$ under the assumption that the ``disturbance" $b(t)$ is bounded \cite{DUetal}, or its derivative is bounded  \cite{TEXZURSCH}. Due to the use of high-gain, the estimation of $i_{dq}(t) $ in both observers occurs in {\em finite time}. There are several drawbacks to this approach: in the observer of  \cite{DUetal} the fault detection depends on the initial conditions of the observer. On the other hand, for the observer of  \cite{TEXZURSCH} its performance depends on the fault severity. An additional drawback of these works is that the information they provide of the ITSCF is ``indirect", since they estimate a signal that is related to the information of interest---namely, the fault current---through the action of an unknown matrix. \\  

\noindent[{\bf W2}] The derivation of an ``approximate" linear time-varying system perturbed by unknown noise to which the authors apply a classical  Kalman filter, artificially attaching a stochastic nature to the unknown noise \cite{zuo2022online,belkhadir2023detection}. Despite its popularity, the Kalman filter has well-known severe  limitations, on one hand, its convergence proof relies on a very strong excitation assumption, encrypted in the assumption of {\em uniform complete observability.} On the other hand, the performance is critically dependent on the system parameters---that being the model a linear {\em approximation} of a highly nonlinear system, are inexact and time-varying.\\

To overcome the above limitations, we propose in this paper two solutions to the problem of on-line detection of ITSCFs in  PMSMs that, unlike the high-gain observers described above, {\em fully exploit} the structure of the ITSCF model.\\

\noindent[{\bf S1}]  A very simple {\em linear observer} with exponential convergence rate. Although this observer guarantees the detection of the fault, the {\em rate of convergence} is fully determined by the motor parameters that, in some cases, may be too slow.\\

\noindent[{\bf S2}] To overcome the latter problem we propose a second solution based on the {\em generalized parameter estimation based observer} (GPEBO) theory \cite{ORTetalaut}. In this design a {\em least squares (LS) estimator} is added to the previous observer to ensure {\em finite convergence time} (FCT). To keep the alertness of the estimator that makes it able to track time-varying parameters a forgetting factor is incorporated in the estimator. Moreover, the FCT is  ensured under the weakest assumption of {\em interval excitation} (IE) of the regressor \cite{KRERIE,TAObook}---the convergence time being exactly equal to the time window of the IE.\\

It should be underscored that, in contrast with existing observers (including the Kalman filter) that provide {\em indirect} information of the fault current, the two latter observers provide {\em explicit} one---namely the amplitude of the fault current.
	
For the case of non-isotropic PMSMs, we develop a parameter estimator that estimates {\em on-line} the resistance and inductance of the {\em faulty} motor. In this way, invoking the ad-hoc {\em certainty equivalent} principle, {\em adaptive} versions of both observers are feasible replacing the {motor's} resistance and inductance of the {\em faulty motor} by their on-line estimated values. To estimate the motor parameters, we derive a regression equation---unfortunately, nonlinear---for the faulty motor. To the best of our knowledge, this is the {\em only} estimation scheme that has been reported for the faulty motor.   Due to the fact that the regression equation that we derive is {\em nonlinear} it is necessary to utilize the Least-Squares + Dynamic Regressor Extension and Mixing (LS+DREM)  parameter estimator of \cite{ORTROMARA}---that, as far as we are aware, is the only parameter estimator that can deal with {\em nonlinear regression equations} (NLRE). Besides this key feature, the LS+DREM estimator is known to be robust to additive noise \cite[Proposition 4]{ORTROMARA} and ensures parameter convergence imposing only an IE assumption \cite{KRERIE,TAObook} on the regressor.
	
	The paper's organization is as follows: The mathematical model of the PMSM, including electrical faults, is presented in Section~\ref{sec2}. A linear observer for detecting ITSCFs in PMSMs is introduced in Section~\ref{sec3}. In Section \ref{sec4} we present the second observer that, in contrast to the linear observer,  ensures FCT.  In Section \ref{sec5} we present a  parameter estimator that is used to render the observers adaptive. {In Section \ref{sec6} we compare the performance of both observers---the linear observer and the GPEBO---with realistic simulation studies.} In Section \ref{sec7} we include some concluding remarks. {In Appendix \ref{appa} we give a list of the acronyms used in the paper. In Appendix \ref{appb} we derive a key state equation for our developments. Appendices \ref{appc} and \ref{appd} are devoted to presenting the parameter estimators, which are used {in establishing Propositions \ref{pro2} and \ref{pro3}, respectively.} In Appendix \ref{appe} we derive the explicit equations of the system model needed for its practical simulation.} 

\section{Fault Model and Problem Formulation}
\label{sec2}
%
	The electrical fault in the faulty winding is modeled with the resistance $R_f \in \rea$, where $i_f(t) \in \rea$ is the current in the short-circuited terms (see, Fig. 1). The intensity of the fault is measured by a scalar $\eta \in [0,1]$, which is the ratio of the number of short-circuited turns to the total number of turns in the faulty phase \cite{DUetal,ZHAetal,CHEetal,TEXZURSCH}. 
	\begin{figure}[htb]
		\centering
		\includegraphics[width=5.5cm,height=4cm]{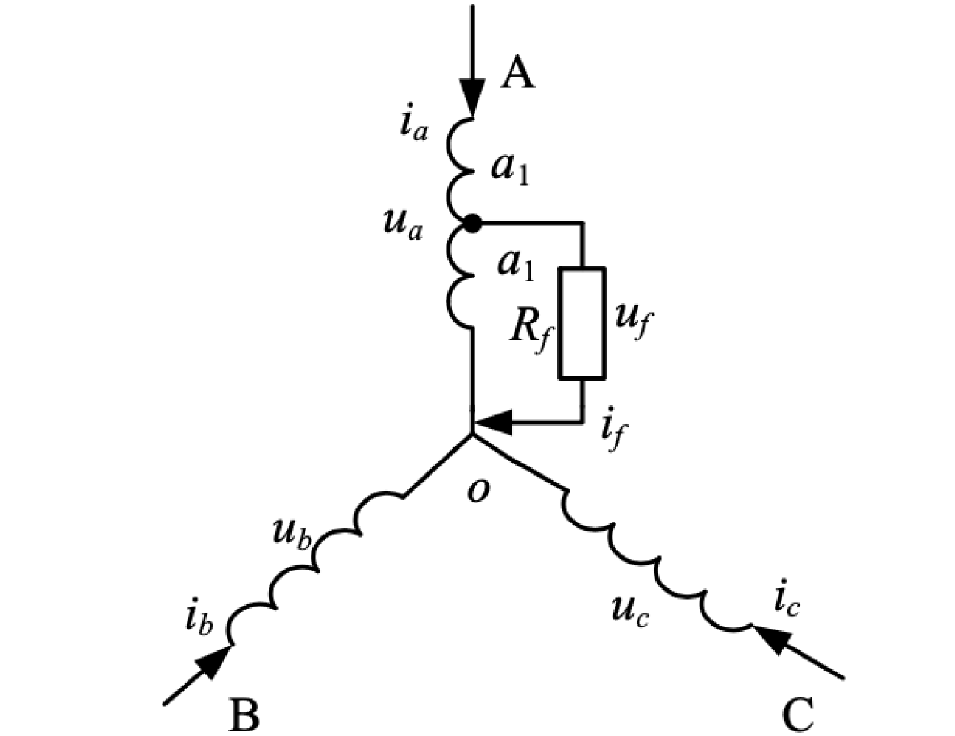}
		\caption{Inter-turn short circuit in phase A for a PMSM.}
		\label{fig1}
	\end{figure}
%
\subsection{Model of the faulty motor}
\label{subsec21}	
	Consider  an interior PMSM in which an ITSCF occurs in phase $A$ as shown in Fig. 1. The $dq$ model dynamical equations are given in \cite[Eq. (3) and (4)]{DUetal}, see also \cite[ (2), (3) and (4)]{TEXZURSCH}\footnote{The mechanical equation is given for the sake of completeness, but it does not play any role in the subsequent ITSCF analysis.}:  
\begsubequ
\lab{sys}
\begali{
\lab{dotid}
{d i_d \over dt} &= {1 \over L_d }(-R_s i_d+\omega L_q i_q + v_d+e_d),\\
\lab{dotiq}
{d i_q \over dt} &= {1 \over L_d }(-R_s i_q-\omega L_d i_d - \omega \phi + v_q+e_q),\\
J\dot \omega &=n_p[(L_d-L_q)i_di_q+\phi i_q]-\tau_L, \\
\lab{ed}
e_d &= {2  \eta\over 3}[(-R_s \cos \theta + \omega(L_d-L_q)\sin \theta) i_f - L_d \cos \theta {d i_f \over dt}],\\
\lab{eq}
e_q &= {2 \eta\over 3} [(R_s \sin \theta - \omega(L_d-L_q)\cos \theta) i_f + L_q \sin \theta {d i_f \over dt}  ],
}
\endsubequ
where $i_d(t)\in\rea$ is the direct current, $i_q(t)\in\rea$ is the quadrature current, {$i_f(t)\in\rea$ is the short-circuit current}, $\omega(t) \in \rea$ is the electrical angular velocity of the machine, $v_d(t)\in\rea$ and $v_q(t)\in\rea$ represent the input voltages applied to the motor and $\theta(t)\in [0,2\pi)$ is the electrical rotor position of the PMSM.  The motor parameters consist of the stator's electrical resistance $R_s>0$, the inductances $L_d>0$ and $L_q>0$, the number of pole pairs $n_p\in\rea$,  the constant permanent magnet flux $\phi\in \rea$, the moment of inertia $J \in\rea$ and  the load torque $\tau_L\in \rea$, which is assumed constant. 

In the present scenario, ${i}_{dq}(t)\in \rea^2$ is {\em known}, however the signal ${e}_{dq}(t)\in \rea^2$ is {\em unknown}, and  expected to be zero under normal conditions. However, the presence of an ITSCF causes deviations from zero in these components. 

\begrem
\lab{rem1}
As mentioned above, the model \eqref{sys} is discussed in \cite{DUetal,TEXZURSCH} and it is ``used" for the observer design writing  \eqref{dotid} and \eqref{dotiq} as
$$
{d i_{dq} \over dt}=a(t)+Q^{-1}e_{dq},
$$
with $Q:=\diag\{L_d,L_q\}$, and trying to estimate $i_{dq}$ and  $e_{dq}$, assuming known $a(t)$ and treating  $e_{dq}$ as an {\em unknown disturbance}, which is assumed bounded in \cite{DUetal} and of bounded derivative in \cite{TEXZURSCH}. Notice that knowledge of  $e_{dq}$ provides, via \eqref{ed} and \eqref{eq}, {\em indirect} information of the signal of interest $i_f$. This kind of ``approximations" of the system dynamics lies at the core of the so-called {\em model-free control} \cite{FLIJOI} used to design PID-like controllers in \cite{FLIJOI1}.
\endrem
%
\subsection{Fault detection procedure}
\label{subsec22}	
The first problem that we address in the paper is to design an algorithm that would provide information about the presence/absence of a fault, via the reconstruction of the signal $\eta i_f(t)$. In \cite[Section 2.3]{MAZBOSDEA} this signal is called {\em Fault Severity Factor}, and is shown to be proportional to the  residual voltage in PMSMs with partitioned windings, which are very often used in applications. The second problem, is the estimation of the electrical parameters of the motor---which is studied for the case of surface-mount PMSMs, that is, with $L_d=L_q$---with the objective of making our observers {\em adaptive}.\footnote{We restrict ourselves to this case here, because the adaptation algorithm for the general case when $L_d \neq L_q$, is extremely involved.} 

The lemma below is instrumental for the solution of the observer design problem.

\beglem
\lab{lem1}
Define the vector signal
\begali{
\lab{iab}
i_{\alpha\beta} &:=\begmat{i_d+({2  \eta\over 3}\cos \theta) i_f \\ i_q - ({2  \eta\over 3}\sin \theta) i_f}.
} 
\noindent [{\bf (i)}] The signal $i_{\alpha\beta}(t) \in \rea^2$ satisfies the differential equation
\begequ
\lab{ltvsys}
Q {d i_{\alpha\beta} \over dt}=A(t) i_{\alpha\beta}+b(t),
\endequ
where and $A(t)$ and $b(t)$ are {\em known} matrices given by
\begalis{
A(t)&:=\begmat{-R_s & \omega(t) L_q \\ -\omega(t) L_d & -R_s},\\
b(t)&:=\begmat{0 \\ -\omega(t) \phi} + v_{dq}(t).
}
\noindent [{\bf (ii)}] The following relation holds
\begequ
\lab{iabminidq}
|i_{\alpha\beta}(t)-i_{dq}(t)| ={2 \eta\over 3} |i_f(t)|.
\endequ
\endlem

\begproo
The details of the derivation of the model \eqref{ltvsys} are given in Appendix \ref{appb}. The proof of \eqref{iabminidq} follows by direct substitution of \eqref{iab} and some simple algebraic calculations.
\endproo

In view of \eqref{iabminidq}, it is clear that {\em reconstructing} the state $i_{\alpha\beta}(t) \in \rea^2$ of system \eqref{ltvsys} provides a solution to the problem of fault detection. In the sequel we give two globally convergent observers of $i_{\alpha\beta}$. The first one is a simple {\em linear, exponentially convergent} observer while the second one is a {\em GPEBO  with FCT} \cite{ORTetalaut}, that relies on the estimation of the systems initial conditions.
%
\section{A Linear Observer-based Solution}
\lab{sec3}
%
\begpro
\label{pro1}
Consider the $dq$ model of an Interior PMSM given in \eqref{sys}, in which an ITSCF occurs in one of the phases. Define the observer
\begequ
\lab{ltvobs}
Q {d \widehat{i_{\alpha\beta}} \over dt}=A(t) \widehat{i_{\alpha\beta}}+b(t).
\endequ

\noindent [{\bf (i)}] The following relation holds
\begequ
\lab{expcon}
|\widehat{i_{\alpha\beta}}(t)-i_{\alpha\beta}(t)|^2 \leq m e^{-\rho t}|\widehat{i_{\alpha\beta}}(0)-i_{\alpha\beta}(0)|^2,
\endequ
where 
\begequ
\lab{mrho}
m:={\max\{{L_d \over L_q},{L_q \over L_d}\} \over \min\{{L_d \over L_q},{L_q \over L_d}\}},\;\rho:=R_s \min\Big\{{1 \over L_q},{1 \over L_d}\Big\}.
\endequ
where the signal $i_{\alpha\beta}$ is defined in \eqref{iab}.

\noindent [{\bf (ii)}] We have that
$$
\liminf |\widehat {i_{\alpha\beta}}(t)-i_{dq}(t)| ={2 \eta\over 3} |i_f(t)|,
$$
exponentially fast.
\endpro

\begproo
To prove the claim [{\bf (i)}], define the observer error signal 
\begequ
\lab{e}
\epsilon_{\alpha\beta}:=\widehat{i_{\alpha\beta}}-i_{\alpha\beta},
\endequ 
which clearly satisfies
$$
Q {d \epsilon_{\alpha\beta} \over dt}=A(t) \epsilon_{\alpha\beta}.
$$

Consider the Lyapunov function candidate
$$
V(\epsilon_{\alpha\beta})=\hal \Big({L_d \over L_q} \epsilon^2_{\alpha}+ {L_q \over L_d}\epsilon^2_{\beta} \Big),
$$
whose time derivative is given by
\begalis{
\dot V & =-{R_s \over L_q} \epsilon^2_{\alpha}- {R_s \over L_d}\epsilon^2_{\beta} \leq -\rho V
}
where $\rho>0$ is defined in \eqref{mrho}. From the inequality above we conclude that
$$
V(t) \leq e^{-\rho t} V(0).
$$
The proof is completed doing some simple bounding on the equation above.

The proof of the claim [{\bf (ii)}] follows evaluating \eqref{iabminidq} with $\widehat {i_{\alpha\beta}}$ and taking into account \eqref{expcon}.
\endproo
%

\section{A Generalized Parameter Estimation-based Observer Solution}
\lab{sec4}
%
The main drawback of the linear observer \eqref{ltvobs} is that there are no degrees of freedom for the improvement of the {\em convergence speed} that---as seen in \eqref{expcon}, \eqref{mrho}---is fully determined by the motor parameters that may be too small to be of practical interest. In this section we present an alternative observer based on the GPEBO theory \cite{ORTetalaut} that overcomes this problem, ensuring FCT. The prize paid for these additional degree of freedom is in the complexity of the GPEBO, that includes several LTI filters and the FCT LS parameter estimator of \cite{ORTetalaut25} that we give in Appendix \ref{appc}.

To simplify the notation we fix a constant $\lambda>0$ and introduce the LTI filter
\begequ
\lab{filf}
\calf(p):={\lambda \over p+\lambda},
\endequ
where $p:={d \over dt}$. The action of this filter on a signal $r(t)$ is denoted as $\calf[r]$ where, for brevity, we omit the time argument. 
\subsection{Proposed GPEBO}
\lab{subsec41}
%
\begpro
\label{pro2}
Consider the filtered signals\footnote{Notice that the filter $p\calf(p)={\lambda p \over p+\lambda}$, hence it is proper.}
\begali{
\nonumber
y_1&:= \calf[R_s\hat{i}_\alpha -v_d-L_q\omega \hat{i}_\beta  ]\\
\nonumber
y_2&:= L_d  p\calf [\hat{i}_\alpha ]\\
\nonumber
\Psi_1 &=\calf[R_s \Phi_1 - L_q\omega \Phi_2]\\
\label{yipsii}
\Psi_2 &:= L_d p \calf[\Phi_1],
}
where $\widehat {i_{\alpha\beta}}(t) \in \rea^2$ is generated via \eqref{ltvobs} and $\Phi_1(t) \in \rea^2$, $\Phi_2(t) \in \rea^2$ are vectors defined as
\begequ
\lab{parphi}
\Phi:=\begmat{\Phi^\top_1 \\ \Phi^\top_2}
\endequ
where $\Phi(t) \in \rea^{2 \times 2}$ is the solution of the matrix equation
\begequ
\lab{dotphi}
\dot \Phi=\mathcal{A}(t) \Phi,\;\Phi(0)=I_2
\endequ
with $\mathcal{A}(t)=Q^{-1}A(t)$.

The following facts hold true.
\begenu
\item[\textbf{F1}] The signal $i_{\alpha\beta}(t) \in \rea^2$, defined in \eqref{iab}, satisfies the equation 
\begin{equation}
\label{iabrep}
i_{\alpha\beta}(t)= \widehat{i_{\alpha\beta}}(t)-\Phi(t) \Theta,
\end{equation}
where $\Theta \in \rea^{2}$ is an {\em unknown} constant vector. 
\item[\textbf{F2}] The vector $\Theta$ verifies the {\em linear regression equation} (LRE)
\begin{equation}
\label{lre}
y(t)= \Psi^\top(t) \Theta,
\end{equation}
where $y(t) \in \rea$ and $\Psi(t) \in \rea^2$ are defined as
$$
y:=\sum_{i=1}^2 y_i,\;\Psi:=\sum_{i=1}^2 \Psi_i,
$$
with $y_i$ and $\Psi_i$ defined in \eqref{yipsii}.
\item[\textbf{F3}]  Let $\Theta_{FCT}(t) \in \rea^{2}$ be an on-line estimate generated with the FCT LS algorithm given in \eqref{thefct} of Proposition \ref{pro4} of Appendix \ref{appc} with the LRE \eqref{lre}. Assume the vector $\Psi(t)$ is IE \cite{KRERIE,TAObook}. That is, there exists $T_c>0$ and $\kappa>0$ such that
\begequ
\lab{ie}
\int_0^{T_c} \Psi(s)\Psi^\top(s)ds \geq \kappa I_2.
\endequ
\endenu
Under these conditions, we have that
\begequ
\lab{concon}
\widehat{i_{\alpha\beta}}(t)-\Phi(t) \Theta_{FCT}(t) = i_{\alpha\beta}(t),\;\forall t \geq T_c,
\endequ
where $\Theta_{FCT}$ is defined in \eqref{thefct}.
\endpro

\begproo
Following the GPEBO procedure consider the error signal \eqref{e}, which clearly satisfies $\dot \varepsilon_{\alpha\beta}=\mathcal{A}(t)\varepsilon_{\alpha\beta} $. In view of \eqref{dotphi} we have that $\varepsilon_{\alpha\beta} (t)=\Phi(t) \Theta$, were we defined $\Theta:=\varepsilon_{\alpha\beta} (0)$, which is a vector of {\em unknown} parameters. From \eqref{e} and the last equation we get \eqref{iabrep}, which proves {\bf F1}.

Replacing \eqref{iabrep} in \eqref{iab}, and taking into account the partition \eqref{parphi}, we get 
\begsubequ
\lab{cossin}
\begali{
\lab{cos}
{2 \eta\over 3} \cos \theta (i_f) &= \hat{i}_\alpha  - \Phi_1^\top \Theta -i_d,\\
\lab{sin}
{2 \eta\over 3} \sin \theta (i_f) &= -\hat{i}_\beta   + \Phi_2^\top \Theta +i_q.
}
\endsubequ
Differentiating \eqref{cos} with respect to time we get 
\begalis{
{2 \eta\over 3} \cos \theta \Big( {d i_{f} \over dt} \Big)&= \omega {2 \eta\over 3} \sin \theta (i_f) + {d \hat{i}_\alpha  \over dt} - \dot{\Phi}_1^\top \Theta - {d i_{d} \over dt} \\
&=  \omega (-\hat{i}_\beta   +i_q + \Phi_2^\top \Theta) + {d \hat{i}_\alpha  \over dt}- \dot{\Phi}_1^\top \Theta - {d i_{d} \over dt}
}
where, to obtain the second equation, we used \eqref{sin}.

Now, let us consider the expression of $e_d$ given in \eqref{sys}:
\begali{
\nonumber
e_d =& -R_s {2 \eta\over 3} \cos \theta (i_f) + \omega(L_d-L_q) {2 \eta\over 3} \sin \theta (i_f)\\ \nonumber
& - L_d {2 \eta\over 3} \cos \theta  {d i_{f} \over dt}\\
\nonumber
=& -R_s (\hat{i}_\alpha  - i_d - \Phi_1^\top \Theta) + \omega(L_d-L_q) (-\hat{i}_\beta   +i_q + \Phi_2^\top \Theta )  \\
&- L_d \Big[\omega (-\hat{i}_\beta   +i_q + \Phi_2^\top \Theta) +  {d \hat{i}_\alpha  \over dt}- {d i_{d} \over dt} - \dot{\Phi}_1^\top \Theta \Big] \nonumber\\
=& -R_s (\hat{i}_\alpha  - i_d - \Phi_1^\top \Theta) -  L_q\omega (i_q - \hat{i}_\beta  + \Phi_2^\top \Theta)  \nonumber \\
&- L_d \Big( {d \hat{i} _\alpha \over dt}- {d i_{d} \over dt} - \dot{\Phi}_1^\top \Theta \Big). 
\label{ed1}
}
where, to obtain the second equation, we used \eqref{cossin} and the equation above.

Let us consider now the first equation in \eqref{sys}, that we repeat here for ease of reference,
$$
L_d {d i_d \over dt} = -R_s i_d+\omega L_q i_q + v_d+e_d.
$$
Applying the filter $\calf(p)$ we get
\begalis{
L_d p \calf[i_d]=& \calf[-R_s i_d+\omega L_q i_q + v_d]+ \calf[e_d]  \\
=& \calf[-R_s i_d+\omega L_q i_q + v_d]-R_s \calf[\hat{i}_\alpha  - i_d - \Phi_1^\top \Theta] \nonumber \\
&-  L_q\calf[\omega (i_q - \hat{i}_\beta  + \Phi_2^\top \Theta)]\nonumber \\
&- L_d p\calf[\hat{i}_\alpha -i_{d}  - \Phi_1^\top \Theta]
}
where we have replaced $e_d$ given in \eqref{ed1} to obtain the second equation. Grouping on the left-hand side the terms independent of  $\Theta$ and on the right-hand side the ones that depend on $\Theta$ we get
\begalis{
&\calf[R_s\hat{i}_\alpha -v_d-L_q\omega \hat{i}_\beta  ]+L_d  p\calf [\hat{i}_\alpha ] \nonumber \\
&=\{\calf[R_s \Phi_1 - L_q\omega \Phi_2] + L_d p \calf[{\Phi}_1]\}^\top \Theta.
}
Invoking the definitions given in the Proposition we recover the LRE \eqref{lre}, which proves {\bf F2}. 

The proof of {\bf F3} follows from Proposition \ref{pro4} in Appendix \ref{appc}, whose proof may be found in \cite{ORTetalaut25,TAObook}. 
\endproo

\begrem
\lab{rem2}
From the fact that $\liminf |\widehat{i_{\alpha\beta}}(t)-i_{\alpha\beta}(t)|=0$ and $i_{\alpha\beta}(t)= \widehat{i_{\alpha\beta}}(t)-\Phi(t) \Theta,\;\forall t \geq 0$, it is clear that $\liminf |\Phi(t)|=0$. This is, indeed, the case because the system \eqref{dotphi} is uniformly asymptotically stable. The relation \eqref{iabrep} is introduced to achieve FCT.
\endrem 
\subsection{Discussion}
\lab{subsec42}
%
As indicated above the main motivation to include a second observer is to give the possibility to adjust the convergence rate of the state estimation and, in particular, in contrast with the linear observer of Proposition \ref{pro1}, to make it independent of the motor parameters. This objective is achieved with the GPEBO observer of Proposition \ref{pro2}, which has FCT, under the IE assumption \eqref{ie}. 

Although there are several parameter estimators that ensure FCT imposing only IE conditions,  {\em e.g.}, \cite{KORetal,ORTNIKGER}, we have selected  one based on LS for the following reasons: 
\begite
\item[(i)] it is robust to additive noise \cite{ORTROMARA}; 
\item[(ii)] {it} has a well-known and simple implementation. Moreover, since it has only one tuning gain ($\gamma$) with a clear practical interpretation, it is easy to tune \cite{SLOLIbook}; 
\item[(iii)] it is based on LS estimation that has better properties than simpler gradient search \cite{RAOTOUbook}; 
\item[(iv)] it incorporates an easily tunable forgetting factor, that allows the algorithm to track time-varying parameters---see \cite{ORTBOBNIK}; 
\endite

Regarding the IE assumption it is important to recall the lemma below, which shows that the IE assumption is {\em necessary and sufficient} to estimate the parameter $\Theta$ with {\em on  or off-line} estimators \cite{GOOSINbook}: 

\begin{lemma} \cite{WANetal}
\lab{lem2}
The LRE \eqref{lre} is {\em identifiable\footnote{See \cite[Definition 2.2]{WANetal} for the definition of ``identifiable" LRE.}} {\em if and only if} the regressor vector $\Psi(t)$ is IE.
\end{lemma}      
%
\section{Estimation of $R_s$ and $L_d(=L_q)$\\ for Isotropic PMSMs}
\lab{sec5}
In this section we consider the case of an isotropic PMSM, {\em e.g.}, $L_d=L_q=:L$, and study the problem of {\em on-line estimation} of the parameters $R_s$ and $L$ for the motor operating in {\em faulty conditions}. This problem is of interest for the implementation of {\em adaptive} versions of the fault detection observers proposed in the previous two sections. That is, we consider the case of certainty-equivalent versions of the observers of Propositions \ref{pro1} and \ref{pro2}, where the parameters $R_s$ and $L$ are replaced by their on-line  estimates. It is reasonable to expect that, if the parameter estimation converges sufficiently fast, the adaptive fault detectors will perform well.

As usual in adaptive parameter estimation problems, the key step for its solution is the derivation of a {\em regression equation}, that relates measurable signals with the unknown parameters. Unfortunately, in the present case, we are unable to derive a {\em LRE}, but we obtain instead a {\em non-linear} one where, moreover, we need to appeal to {\em overparameterization}. As is well-known, overparameterization has a deleterious effect in the performance of the parameter estimator. On the other hand, the fact that the regression is nonlinear is not critical because, as we prove in the following proposition, the regression is {\em ``monotonizable"}, and the LS+DREM estimator of Appendix \ref{appd} is able to handle these kind of parameterizations.       

\begpro
\label{pro3} 
Consider the $dq$ model of an Interior PMSM, in which an ITSC fault occurs in one of the phases, given in \eqref{sys}. 

The following properties hold true.

\begenu
\item[\textbf{P1}]  It is possible to construct a measurable signal $\caly(t) \in \rea$ and a measurable regressor $\xi(t) \in \rea^5$ such that the following NLRE holds
\begequ
\lab{nlre}
\caly  = \xi^\top \calg(\mu),
\endequ
where $\calg:\rea^3 \to \rea^5$ is a {\em ``monotonizable"} map,\footnote{See Definition 1 in Appendix \ref{appd} for the characterization of a ``monotonizable" map.} with $\mu:=\col({R_s \over L},{1 \over L}, \mu_3)$.\footnote{Notice that, besides having to deal with a NLRE, the map $\calg(\mu)$ is {\em overparameterized}, including a parameter $\mu_3$ of no practical interest.}  
\item[\textbf{P2}] Assume the {\em regressor $\xi(t)$ is IE}. 

If the parameters $\mu$ are estimated with the LS+DREM algorithm of Proposition \ref{pro5} in Appendix \ref{appd} we have the following.
\begenu
\item [{\bf (i)}] For all initial conditions the estimated parameters verify
$$
\liminf \hat \mu(t)= \mu,
$$
exponentially fast.
\item [{\bf (ii)}] All the signals are {\it bounded}.
\endenu
\endenu
\endpro

\begproo
For the isotropic PMSM the vector $e_{dq}$ \eqref{ed}, \eqref{eq} takes the form
\begalis{
e_d&=-{2 \eta\over 3} \cos \theta (R_s i_f + L {di_f \over dt})\\
e_q&={2 \eta\over 3} \sin \theta (R_s i_f + L {di_f \over dt})
}
which clearly satisfy
$$ 
\sin \theta e_d + \cos \theta e_q=0.
$$
Combining this identity with the current dynamics \eqref{dotid}, \eqref{dotiq} we get
\begali{
\lab{swal}
\sin \theta  {di_d \over dt} + \cos \theta {di_q \over dt}=&{1 \over L}\Big[\sin \theta (-R_s i_d+\omega L i_q + v_d) \nonumber \\
&+ \cos \theta (-R_s i_q-\omega L i_d - \omega \phi + v_q)\Big].
}
Applying the filter $\calf(p)$ to the equation above we get
\begali{
\lab{swalem}
\calf \Big[ \sin \theta  {di_d \over dt} + \cos \theta  {di_q \over dt}\Big] =& \sin \theta p \calf [ i_d ] - {1 \over \lambda} \calf [ \omega \cos \theta p \calf [ i_d ]] \nonumber \\
&+ \cos \theta p \calf [ i_q ] + {1 \over \lambda} \calf [ \omega \sin \theta p \calf [ i_q ]],
}
where we used the Swapping Lemma\footnote{\cite[Lemma 3.6.5]{SASBODbook} Consider the filter $\calf(p)$ given in \eqref{filf} and the signals $w(t) \in \rea^{n \times m}$ and $v(t) \in \rea^m$. Then:
$
\calf[wv]=\calf[w]v-\calf\Big[{1 \over p+\lambda}[w]\dot v\Big].
$
} 
to obtain the identity. Now, integrating the identity
$$
\dot \theta=\omega,\;\theta(0)=\theta_0,
$$
we get
$$
\theta(t) = \theta_0 + \int_0^{t} \omega (\tau) d\tau =: \theta_0 + v_\omega(t).
$$
Consequently, we have
		\begin{align}
			\sin \theta &= \sin (\theta_0 + v_\omega) \notag \\
			&= \sin \theta_0 \cos v_\omega + \cos \theta_0 \sin v_\omega =: \eta_1 \phi_1 + \eta_2 \phi_2, \notag \\ 
			\cos \theta &= \cos (\theta_0 + v_\omega) \notag \\
			 &= \cos \theta_0 \cos v_\omega - \sin \theta_0 \sin v_\omega =: \eta_2 \phi_1 - \eta_1 \phi_2. 
       \lab{sincos}
		\end{align}
where, in the last identity, we defined
\begalis{
\eta_1 &:=\sin \theta_0,\;\eta_2 :=\cos \theta_0,\\
\phi_1 &:=\cos v_\omega,\;\phi_2 :=\sin v_\omega.
}
Applying the identities \eqref{sincos} to the right hand side of \eqref{swalem} we get
\begali{
\nonumber
&\calf \Big[ \sin \theta  {di_d \over dt} + \cos \theta  {di_q \over dt}\Big] \\
\nonumber
=&  (\eta_1 \phi_1 + \eta_2 \phi_2) p \calf [ i_d ] - {1 \over \lambda} \calf [ \omega (\eta_2 \phi_1 - \eta_1 \phi_2) p \calf [ i_d ]]\\
\nonumber
& +  (\eta_2 \phi_1 - \eta_1 \phi_2) p \calf [ i_q ]+ {1 \over \lambda} \calf [ \omega (\eta_1 \phi_1 + \eta_2 \phi_2) p \calf [ i_q ]] \\
\lab{eta1phi3}
=&: \eta_1 \phi_3 + \eta_2 \phi_4,
}
where to obtain the second identity we group terms and defined
\begalis{
\phi_3:=&\phi_1 p \calf [ i_d ] + {1 \over \lambda} \calf [ \omega  \phi_2 p \calf [ i_d ]] - \phi_2 p \calf[i_q] \\
\nonumber
&+ {1 \over \lambda} \calf [ \omega  \phi_1 p \calf [ i_q ]],\\
\phi_4:=&\phi_2 p \calf [ i_d ]- {1 \over \lambda} \calf [ \omega  \phi_1 p \calf [ i_d ]] + \phi_1 p \calf[i_q]\\
\nonumber
&  + {1 \over \lambda} \calf [ \omega  \phi_2 p \calf [ i_q ]].
}
Applying the identities \eqref{sincos} to {the first term on the right-hand side of \eqref{swal}} we get
\begalis{
&\calf \Big[{\sin \theta \over L }(-R_s i_d+\omega L i_q + v_d)\Big]\\
\nonumber
=& \calf \Big[{1 \over L }(\eta_1 \phi_1 + \eta_2 \phi_2) (-R_s i_d+\omega L i_q +v_d)\Big] \\
=& -\eta_1 {R_s \over L } \calf [\phi_1 i_d] - \eta_2 {R_s \over L } \calf [\phi_2 i_d] + \eta_1 \calf [ \phi_1 \omega i_q] \\
\nonumber
& + \eta_2 \calf [\phi_2 \omega i_q]+ {\eta_1 \over L } \calf [\phi_1 v_d] + {\eta_2 \over L } \calf [\phi_2 v_d]\\
=:& \eta_1 {R_s \over L } \phi_5 + \eta_2 {R_s\over L } \phi_6 + \eta_1 \phi_7 + \eta_2 \phi_8 + {\eta_1 \over L } \phi_9 + {\eta_2 \over L } \phi_{10},
}
where to obtain the second identity we group terms and defined
\begalis{
\phi_5 &:=- \calf [\phi_1 i_d],\;\phi_6 :=-\calf [\phi_2 i_d],\;\phi_7:=\calf [\phi_1 \omega i_q],\\
\phi_8 &:= \calf [\phi_2 \omega i_q],\;\phi_9 :=\calf [\phi_1 v_d],\;\phi_{10}:=\calf [\phi_2 v_d].
}
Applying the identities \eqref{sincos} to {the second term on the right-hand side of \eqref{swal}} we get
\begalis{
&\calf \Big[{\cos \theta \over L } (-R_s i_q-\omega L i_d - \omega \phi + v_q)\Big]\\
\nonumber
=&\calf \Big[ {1\over L }(\eta_2 \phi_1 - \eta_1 \phi_2) (-R_s i_q-\omega L i_d - \omega \phi + v_q)\Big] \\
=& \eta_1 {R_s \over L } \calf [\phi_2 i_q] - \eta_2 {R_s\over L } \calf [\phi_1 i_q] + \eta_1 \calf [\phi_2 \omega i_d] - \eta_2 \calf [\phi_1 \omega i_d] \\
&+ {\eta_1 \over L } \calf [\phi_2 (\omega \phi- v_q)] + {\eta_2 \over L } \calf [\phi_1 (-\omega \phi+ v_q)]\\
=:&\eta_1 { R_s \over L }\phi_{11} + \eta_2 { R_s \over L }\phi_{12} + \eta_1 \phi_{13} + \eta_2 \phi_{14} + {\eta_1 \over L } \phi_{15} + {\eta_2 \over L } \phi_{16},
}
where to obtain the second identity we group terms and defined
\begalis{
\phi_{11} &:=\calf [\phi_2 i_q],\;\phi_{12} :=-\calf [\phi_1 i_q],\;\\ \phi_{13} &:=\calf [\phi_2 \omega i_d],\;\phi_{14}:=-\calf [\phi_1 \omega i_d], \\\phi_{15} &:=\calf [\phi_2 (\omega \phi- v_q)],\;\phi_{16} :=\calf [\phi_1 (-\omega \phi+ v_q)].
}

Further from \eqref{eta1phi3}, we get
\begali{
\nonumber
\eta_1 \phi_3 + \eta_2 \phi_4=& { R_s \over L} (\eta_1  \phi_5 + \eta_2 \phi_6) + \eta_1 \phi_7 + \eta_2 \phi_8 \\
 \nonumber
&+ { 1 \over L} (\eta_1 \phi_9 + \eta_2 \phi_{10}) + { R_s \over L} (\eta_1 \phi_{11} + \eta_2 \phi_{12}) \\
 \nonumber
&+ \eta_1 \phi_{13} + \eta_2 \phi_{14} + { 1 \over L} (\eta_1 \phi_{15} + \eta_2 \phi_{16}).
}
Thereby, it follows that
$$
\eta_1 \caly ={ R_s \over L} (\eta_1 \xi_1 + \eta_2 \xi_2) + { 1 \over L} (\eta_1 \xi_3 + \eta_2 \xi_4)+\eta_2 \xi_5,
$$
where        
\begalis{
\caly &:= \phi_3 - \phi_7 - \phi_{13},\\ 
\xi_1 &:= \phi_5 + \phi_{11}, \\ 
\xi_2 &:= \phi_6 + \phi_{12}, \\
\xi_3 &:= \phi_9 + \phi_{15}, \\
\xi_4 &:= \phi_{10} + \phi_{16}, \\
\xi_5 &:= -\phi_4 + \phi_8 + \phi_{14}. \\ 
}
Dividing by $\eta_1$ and defining
\begalis{
\xi & :=\begmat{\xi_1\\ \xi_2 \\ \xi_3 \\ \xi_4 \\ \xi_5},\; \mu := \begmat{\mu_1 \\ \mu_2 \\ \mu_3}=\begmat{{R_s \over L} \\ {1 \over L} \\ {\eta_2 \over \eta_1}},\;
\calg(\mu) := \begmat{\mu_1\\ \mu_1 \mu_3 \\ \mu_2 \\  \mu_2 \mu_3 \\ \mu_3},
}
we obtain the NLRE \eqref{nlre} completing the proof of claim {\bf P1}. 

The proof of {\bf P2} follows from Proposition \ref{pro5} in Appendix \ref{appd}, whose proof may be found in \cite{ORTROMARA}.
\endproo
%
\section{Simulation Results}
\lab{sec6}
%
We consider an isotropic PMSM with an ITSCF in phase A, as described in \eqref{sys}. The motor parameter values are taken from \cite{TEXZURSCH} and are listed in {Table 1.} To track the desired motor speed $\omega = 300~\text{rad/s}$ under a load torque of $10~\text{N} \cdot \text{m}$, the same cascaded PI controller used in \cite{TEXZURSCH} is adopted with identical controller parameter values $k_{p,i_d}=k_{p,i_q}=0.5$, $k_{p,w}=2$, $T_{i,i_d}=T_{i,i_q}=0.0011$ and $T_{i,w}=0.0011$. 
 
 \begin{table}[h]
		\centering
		\caption{Parameter values of PMSM}\label{Tab:motparam}
		\centering
		\renewcommand {\arraystretch}{1}
		\begin{tabular}{|c|c|c|c|}
			\hline
			Parameters&Description& Value\\ \hline 
			$n_p$&Number of pole pairs& $3$\\  \hline 
			$L_d$&Direct axis inductance&1.679 mH\\\hline 
			$L_q$&Quadrature axis inductance&1.679 mH\\\hline 
			$R_s$&Stator resistance& 1.5 $\Omega$ \\\hline 
			$J$&Moment of inertia&$3.6\times10^{-3}$ kgm$^2$\\\hline 
			$\phi$&Permanent magnet flux  & $0.1725$ \textrm{Wb}\\\hline 
			$R_f$&Fault resistance&$5$ $\Omega$\\
			\hline
		\end{tabular}
	\end{table}

In order to address the problem of online detection of ITSCFs in PMSMs, the open loop observer (OLO) and the GPEBO were considered. The OLO was designed following Proposition \ref{pro1}, while the GPEBO is designed according to Proposition \ref{pro2}, with the observer of $i_{\alpha\beta}$ given in \eqref{ltvobs}, and the parameter estimator defined in \eqref{lsd} and \eqref{thefct}. 

In the simulation, the observer parameter values are set to $\lambda=100$, $\gamma=1$, $\chi_0=4$, $k=15$, $f_{11}=1$, $f_{12}={1 \over 5}$, $f_{21}={1 \over 4}$, $f_{22}={1 \over 3}$, \[
G = \begin{bmatrix}
6.25\times10^{5} & 0 \\
0 & 6.25\times10^{12}
\end{bmatrix}.
\]
We use the same initial values of the system states and the observers as those used in \cite{TEXZURSCH}, namely: $i_{dq}(0)=[0,0]^\top$, $w(0)=0$, $i_f(0)=0$, $\widehat{i_{\alpha\beta}}(0)=[1,1]^\top$. According to the definition of $\Theta:=\varepsilon_{\alpha\beta} (0)$, the unknown parameter vector is identified as $\Theta=\widehat{i_{\alpha\beta}}(0)-i_{\alpha\beta}(0)=[1,1]^\top$. The initial estimate of the parameter vector is set to  $\hat{\Theta}(0)=[0,0]^\top$. A short circuit, with $\eta=0.4$, is introduced at $t=0.1$ seconds.

 The simulation results are presented in Fig. 2 to Fig. 6. The fault manifests as oscillations in the currents as shown in Figs. 2 and 3. As shown in Figs. 2 and 3, both observers can accurately estimate the system states, with the GPEBO exhibiting a faster convergence compared to the OLO. Fig. 4 shows the norm of the estimated fault current and the norm of the actual fault current, revealing that they are almost indistinguishable in a very short time. In Figs. 5 and 6 we show the transient behavior of the parameters {$\hat \Theta$}, it can be seen that, the proposed algorithm ensures that the estimates of the unknown parameters converge to their true values in a finite, very short, time.

\begin{figure} 
\begin{centering}
\includegraphics[width=\columnwidth]{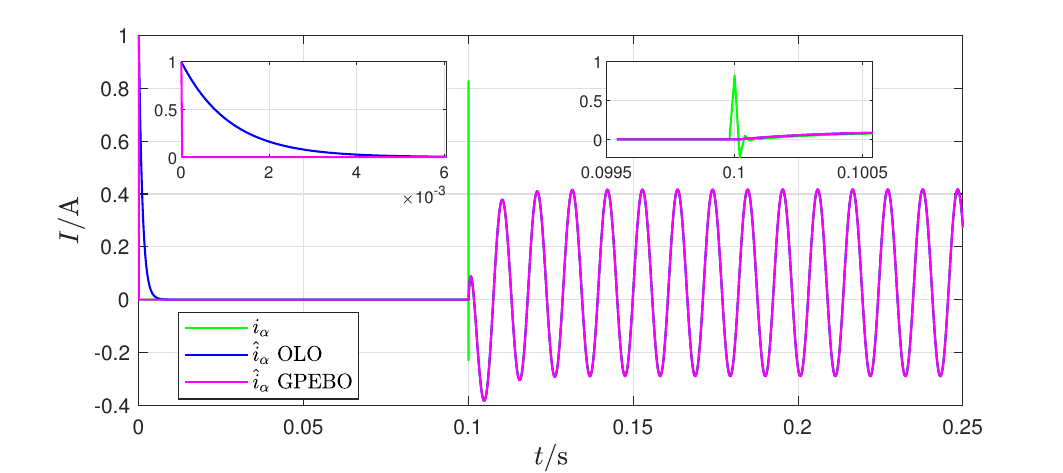}
\par\end{centering}
\caption{\label{fig2} Comparison of the current $i_\alpha $ and its estimate $\hat{i}_\alpha $ using OLO and GPEBO}
\end{figure}

\begin{figure}
\begin{centering}
\includegraphics[width=\columnwidth]{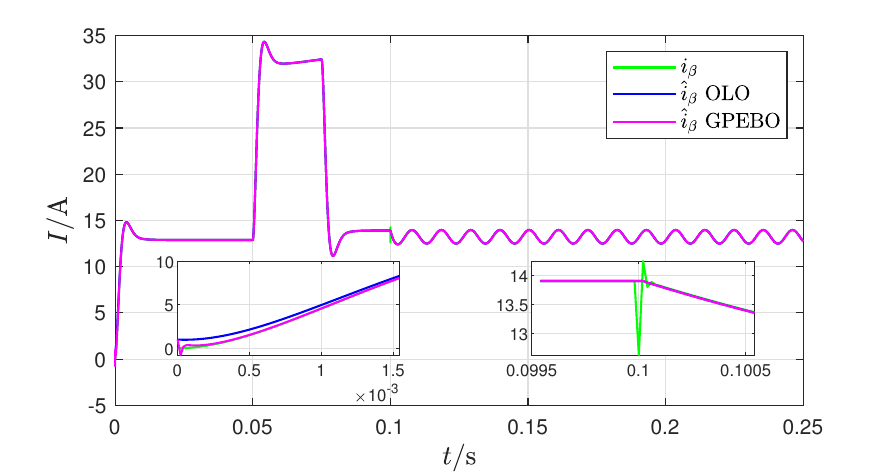}
\par\end{centering}
\caption{\label{fig3} Comparison of the current $i_\beta  $ and its estimate $\hat{i}_\beta  $ using OLO and GPEBO}
\end{figure}

\begin{figure}
\begin{centering}
\includegraphics[width=\columnwidth]{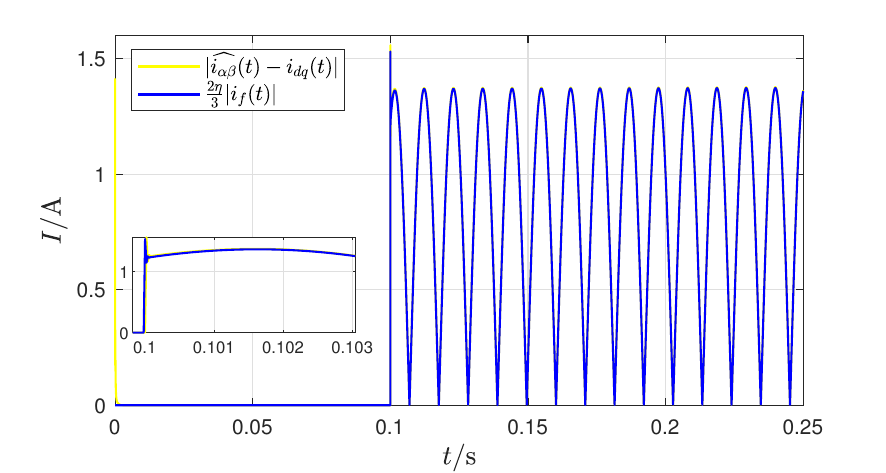}
\par\end{centering}
\caption{\label{fig6} The norm of the {\em estimated} fault current $|\widehat{ i_{\alpha\beta}}(t)-i_{dq}(t)|$ and the norm of the {\em actual} fault current ${2 \eta\over 3} |i_f(t)|$. }
\end{figure}

\begin{figure}
\begin{centering}
\includegraphics[width=\columnwidth]{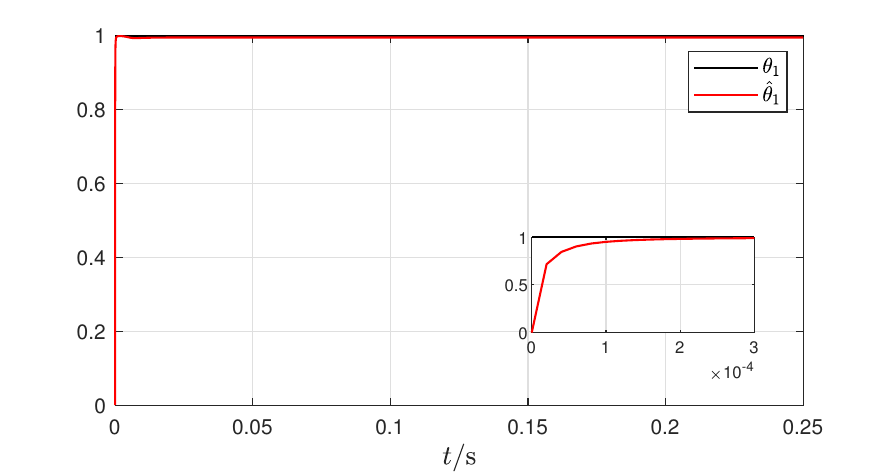}
\par\end{centering}
\caption{\label{fig4} The estimated value $\hat{\theta}_1$ and the actual one  ${\theta}_1$ of GPEBO}
\end{figure}

\begin{figure}
\begin{centering}
\includegraphics[width=\columnwidth]{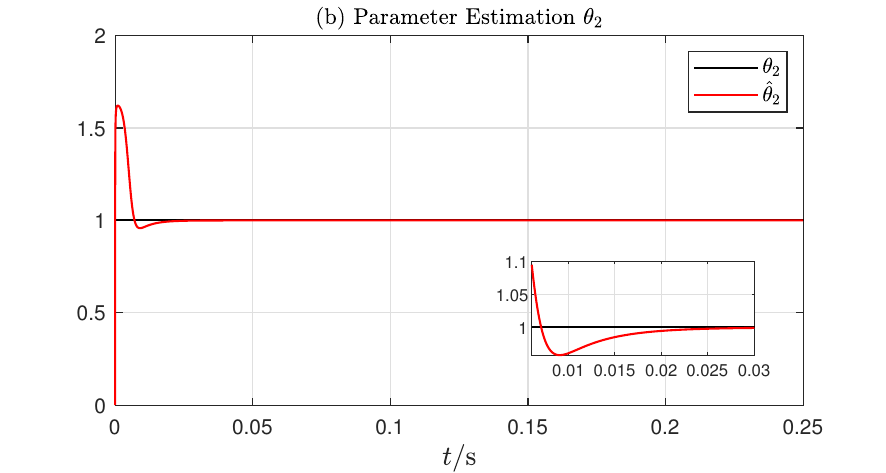}
\par\end{centering}
\caption{\label{fig5} The estimated value $\hat{\theta}_2$  and the actual one  ${\theta}_2$ of GPEBO}
\end{figure}

%
\section{Concluding Remarks}
\lab{sec7}
%
We have proposed here two solutions to the problem of detection of ITSCFs in PMSMs for the case when the motor parameters are known. Moreover, for the case when $L_d=L_q$, an adaptive version of these observers that estimate the motor parameters has been developed. Simulation results comparing the performance of both observers are presented.

Our future research is concentrated in the following points:
\begite
\item the comparison of our observers with the standard Kalman filter-based solution to the problem;
\item  the implementation of the adaptive version of the observers, including variations in the motor parameters;
\item the validation of the results in an experimental facility.
\endite    

\appendices
\section{List of Acronyms}
\label{appa}
\renewcommand{\arraystretch}{1.4} 
\begin{center}
\hspace{1cm} 
\fbox{ \begin{minipage} {1\linewidth}
\begin{tabular}{ll}
DREM & Dynamic regressor extension and mixing\\
FCT & Finite convergence time \\
GPEBO & Generalized parameter estimation based observer\\
LS & Least squares\\
IE & Interval excitation\\
IPMSM & Interior permanent magnet synchronous motor\\
ITSCF & Interturn short circuit fault\\
LRE & Linear regression equation\\
LTI & linear time invariant\\
NLRE  & Nonlinear regression equation\\
PMSM & Permanent magnet synchronous motor\\
\end{tabular}
\end{minipage}}
\hspace{1cm}
\end{center}
\section{Mathematical Derivation of System (4)}
\label{appb}
In order to estimate the Fault Severity Factor $\eta i_f(t)$, we reconstruct the state $i_{\alpha\beta}$ of the system \eqref{ltvsys}. The analytical derivation of system \eqref{ltvsys} from \eqref{sys} and \eqref{iab} is detailed below.

Define the rotation matrix 
$$
\calj:=\begmat{0 & -1 \\ 1 & 0},$$ so that 
$$
e^{-\cal\calj \theta}=\begmat{\cos \theta & \sin \theta \\ -\sin \theta  & \cos \theta}.
$$ 
Using this, along with the definition of $e_{dq}=[e_d, e_q]^\top$ in \eqref{sys} and the reconstructed state $i_{\alpha\beta}$ in \eqref{iab}, we can obtain
\begsubequ
\lab{eij}
\begali{
\lab{edqj}
 e_{dq}=&{2\eta \over 3}\begmat{-R_s & -\omega(L_d-L_q) \\ -\omega(L_d-L_q) & -R_s} e^{-\calj \theta}{\bf e}_1^\top i_f \nonumber \\
&-{2\eta \over 3}Q e^{-\calj \theta}{\bf e}_1^\top \idot_f ,\\
 \lab{iabj}
 i_{\alpha\beta}&=i_{dq}+{2\eta \over 3}e^{-\calj \theta}{\bf e}_1^\top i_f,
}
\endsubequ
where ${{\bf e}_1:=[1, 0]}.$
Differentiating {$Q i_{\alpha\beta}$} with respect to time yields
\begali{
\lab{diabj}
Q\idot_{\alpha\beta}=& Q\idot_{dq} + {2\eta \over 3}\begmat{0 & L_d\omega \\-L_q\omega & 0} e^{-\calj \theta}{\bf e}_1^\top i_f \nonumber \\
&+{2\eta \over 3}Q e^{-\calj \theta}{\bf e}_1^\top \idot_f.
}
Since
$$
Q\idot_{dq}=\begmat{-R_s & \omega L_q \\ -\omega L_d & -R_s}i_{dq}+\begmat{0 \\ -\omega\phi}+v_{dq}+e_{dq},
$$
substituting this and \eqref{eij} into \eqref{diabj} gives
$$
\begin{aligned}
Q\idot_{\alpha\beta}=&\begmat{-R_s & \omega L_q \\ -\omega L_d & -R_s}i_{dq}+\begmat{0 \\ -\omega\phi}+v_{dq} \nonumber \\
&+{2\eta \over 3}\begmat{-R_s & \omega L_q \\ -\omega L_d & -R_s} e^{-\calj \theta}{\bf e}_1^\top i_f\\
=&\begmat{-R_s & \omega L_q \\ -\omega L_d & -R_s}\Big( i_{dq}+{2\eta \over 3} e^{-\calj \theta}{\bf e}_1^\top i_f\Big)\nonumber \\
&+\begmat{0 \\ -\omega\phi}+v_{dq}\\
=&\begmat{-R_s & \omega L_q \\ -\omega L_d & -R_s}i_{\alpha\beta}+\begmat{0 \\ -\omega\phi}+v_{dq},
\end{aligned}
$$
which is of the form \eqref{ltvsys}.

\section{A Least Squares Estimator \\with Finite Convergence Time}
\label{appc}
%
Surprisingly, it is not well-known that, for a LRE, the standard LS estimator (with or without forgetting factor) {\em ensures FCT} under the assumption of IE of the regressor. The result is contained in \cite[Lemma 3.9]{TAObook} for LS without forgetting factor and, for LS with forgetting factor, it follows directly from the proof of \cite[Proposition 1]{ORTROMARA} and was recently reported in \cite{ORTetalaut25}. For the sake of completeness we repeat this result here, whose proof follows immediately from the proof of \cite[Proposition 1]{ORTROMARA}, where it is shown that the matrix $I_2-z(t)f_0F(t)$ is full rank for $t \geq T_c$. \\
 
\begpro
\lab{pro4}
Consider the regression equation \eqref{lre} and assume $\Psi(t)$  is IE and {\it bounded}.  Define the standard LS estimator with forgetting factor \cite[Subsection 8.7.6]{SLOLIbook}:\footnote{To improve the condition number of the matrix $F$, we have added a matrix $G$ to the estimator of \cite[Subsection 8.7.6]{SLOLIbook}, which does not affect the proof of the claim.}
\begsubequ
\lab{lsd}
\begali{
\lab{lsd1}
		\dot{\hat \Theta} & =\gamma FG\Psi (y-\Psi^\top \hat\Theta),\; \hat\Theta(0)=\Theta_{0} \in \rea^{2},\\
\lab{lsd2}
\dot {F}& =  -\gamma F G \Psi \Psi^\top F +\chi F,\; F(0)=F_0=\begmat{f_{11} & f_{12}\\ f_{21} & f_{22}} \in \rea^{2\times2},\\
\lab{lsd3}
\dot z &=\; -\chi z, \; z(0)=1, \\
\chi &= \chi_0 \left( 1-{{\| F\|}\over{k}} \right),\;k\geq {\| F_0\|}
}
\endsubequ
with tuning gain the scalar $\gamma>0$, $f_{11}>0$, $f_{12}>0$, $f_{21}>0$, $f_{22}>0$, $\chi_0> 0$ and $k\geq  {\| F_0\|}$, and the matrix $G>0$.

For $t \geq T_c$, define the signal
\begequ
\lab{thefct}
\Theta_{FCT}(t):=[I_2-z(t)F(t)F^{-1}_0]^{-1}[\hat \Theta(t) - z(t)F(t)F^{-1}_0\Theta_0].
\endequ
\begenu
\item [{\bf (i)}] 
For all initial conditions this signal verifies
$$
\Theta_{FCT}(t)= \Theta,\;\forall t \geq T_c.
$$
\item [{\bf (ii)}] The identity \eqref{concon} is satisfied.
\item [{\bf (iii)}] All the signals are {\it bounded}.
\endenu 
\endpro
%
\section{The LS+DREM Parameter Estimator of \cite{ORTROMARA}}
\label{appd}
%
\begin{definition}
\lab{def1}
A mapping $\calg:\rea^3 \to \rea^5$ is said to ``monotonizable" if there exists a matrix  $P\in\mathbb{R}^{3\times 5}$ such that mapping $\calg(\mu)$ verifies the linear matrix inequality
\begalis{
	P\nabla\calg(\mu) + \nabla^\top  \calg(\mu)P^\top  \geq \rho I_3 >  0,\;\forall \; \mu \in \rea^3,
}
for some $\rho \in \rea_{>0}$. Consequently  \cite{DEM}, the mapping {$\calg(\mu)$} is {\em strongly monotone}, that is, 
\begalis{
(a-b)^\top \left[P \calg(a) -P \calg(b)\right] &\geq \rho|a-b|^2  >  0,\; \nonumber\\
\forall \; a,b \in \rea^q,\;a  &\neq b.
}
\end{definition}

The main properties of the LS+DREM estimator are summarized in the proposition below. The only step that needs to be proven is the mapping $\calg(\mu)$ is ``monotonizable", with the remaining part of the proof found in \cite{ORTROMARA}. This is easy to verify with the matrix
$$
P=\begmat{I_3 & 0_{3 \times 2}},
$$
for which we get $P\nabla\calg(\mu)=I_3$. 

\begpro
\lab{pro5}
Consider the NLRE \eqref{nlre} and assume {$\xi(t)$}  is IE and {\it bounded}.  Define the LS+DREM estimator with forgetting factor \cite{ORTROMARA} 
\begalis{
		\dot{\hat \calg} & =\gamma_\calg F {\xi} (y-{\xi^\top} \hat\calg),\; \hat\calg(0)=\calg_{0} \in \rea^{5}\\
\dot {F}& =  -\gamma_\calg F {\xi\xi^\top} F +\chi F,\; F(0)={1 \over f_0} I_{5} \\
\dot{\hat \mu} & = {\gamma_{\mu}}  \Delta P [{\bf Y} -\Delta \hat\mu ],\; \hat\mu(0)=\mu_{0} \in \rea^3\\
\dot z &=\; -\chi z, \; z(0)=1, \\
\chi &= \chi_0 \left( 1-{{\| F\|}\over{k}} \right)
}
with tuning gains the scalars $\gamma_\calg>0, {\gamma_{\mu}>0}$, $f_0>0$, $\chi_0> 0$ and $k\geq \frac{1}{f_0}$, and we defined the signals
	\begalis{
		\Delta & :=\det\{I_{5}-z f_0F\}\\
		{\bf Y} & := \adj\{I_{5}- zf_0F\} (\hat\calg -  zf_0F \calg_{0}),
	}
where $ \adj\{\cdot\}$ denotes the adjugate matrix.
\begenu
\item [{\bf (i)}] For all initial conditions the estimated parameters verify
$$
\liminf|\hat \mu(t)- \mu|=0,
$$
exponentially fast.
\item [{\bf (ii)}] All the signals are {\it bounded}.
\endenu 
\endpro
%
\section{Explicit Ordinary Differential Equation Model}
\lab{appe}
Based on \cite{FLetal}, the voltage vector equation of the IPMSM with an ITSC fault expressed in the rotating \emph{dq} coordinate, is given by
 
\begin{equation}
\label{ode}
\begin{aligned}
\begin{bmatrix} v_d  \\ v_q \\ 0 \end{bmatrix} 
&= 
\begin{bmatrix} 
R_s & 0  & \frac{2 \eta}{3} R_s \cos \theta \\ 
0  & R_s  & -\frac{2 \eta}{3} R_s \sin \theta \\ 
\eta R_s \cos \theta  & -\eta R_s \sin \theta  & \eta R_s +  R_f 
\end{bmatrix}
\begin{bmatrix} i_d  \\ i_q \\ i_f \end{bmatrix} \\
&\quad +
\begin{bmatrix} -\omega \psi_q \\ \omega \psi_d \\ 0 \end{bmatrix}
+
\frac{d}{dt} 
\begin{bmatrix} \psi_d \\ \psi_q \\ \psi_{\text{fault}} \end{bmatrix}
\end{aligned}
\end{equation}
with
\begin{equation}
\begin{aligned}
\begin{bmatrix}
\psi_d \\
\psi_q \\
\psi_{\text{fault}}
\end{bmatrix}
&=
\begin{bmatrix}
L_d & 0 & \frac{2 \eta}{3} L_d \cos \theta \\
0 & L_q & -\frac{2 \eta}{3} L_q \sin \theta \\
\eta L_d \cos \theta & -\eta L_q \sin \theta & \frac{2}{3} \eta^2 L_{AA}
\end{bmatrix}
\begin{bmatrix}
i_d \\
i_q \\
i_f
\end{bmatrix}
\\
& \quad + \begin{bmatrix}
\phi \\
0 \\
\eta \phi \cos \theta
\end{bmatrix}
\end{aligned}
\end{equation}
where {$R_f > 0$ is the fault resisttance,} $L_{AA}$ is the self-inductance of phase A. Taking the time derivative of $[\psi_d, \psi_q, \psi_{\text{fault}}]^\top$ yields
\begequ
\lab{dpsi}
\begin{aligned}
\begin{bmatrix} 
\dot{\psi}_d \\ 
\dot{\psi}_q \\ 
\dot{\psi}_{\text{fault}} 
\end{bmatrix}
=&
\begin{bmatrix} 
0 & 0  & -\frac{2 \eta}{3} L_d \omega \sin \theta \\ 
0  & 0  & -\frac{2 \eta}{3} L_q \omega \cos \theta \\ 
-\eta L_d \omega \sin \theta  & -\eta L_q \omega \cos \theta  & \frac{2}{3}\eta^2 \dot{L}_{AA} 
\end{bmatrix}
\begin{bmatrix} 
i_d  \\ 
i_q \\ 
i_f 
\end{bmatrix}
\\
&+ \ 
\begin{bmatrix} 
L_d & 0  & \frac{2 \eta}{3} L_d \cos \theta \\ 
0  & L_q  & -\frac{2 \eta}{3} L_q \sin \theta \\ 
\eta L_d \cos \theta  & -\eta L_q \sin \theta  & \frac{2}{3}\eta^2 L_{AA} 
\end{bmatrix}
\begin{bmatrix} 
\frac{d}{dt}i_d  \\ 
\frac{d}{dt}i_q \\ 
\frac{d}{dt}i_f 
\end{bmatrix}\\
&+
\begin{bmatrix} 
0  \\ 
0 \\ 
-\eta \phi \omega \sin \theta
\end{bmatrix}.
\end{aligned}
\endequ
Substituting \eqref{dpsi} into \eqref{ode} we get {
$$
\begin{aligned}
&\begin{bmatrix} 
L_d & 0  & \frac{2 \eta}{3} L_d \cos \theta \\ 
0  & L_q  & -\frac{2 \eta}{3} L_q \sin \theta \\ 
\eta L_d \cos \theta  & -\eta L_q \sin \theta  & \frac{2}{3}\eta^2 L_{AA} 
\end{bmatrix}
\begin{bmatrix} 
\frac{d}{dt}i_d  \\ 
\frac{d}{dt}i_q \\ 
\frac{d}{dt}i_f 
\end{bmatrix}
\\
&= 
-\begin{bmatrix} 
Rs & -\omega L_q  & M_{13}\\
\omega L_d  & Rs  & M_{23}\\ 
M_{31}  & M_{32} & M_{33}
\end{bmatrix}
\begin{bmatrix} 
i_d  \\ 
i_q \\ 
i_f 
\end{bmatrix}
+
\begin{bmatrix} 
v_d  \\ 
v_q - \omega \phi\\ 
\eta \phi \omega \sin \theta 
\end{bmatrix}.
\end{aligned}
$$
where 
\begin{equation}
\left\{
\begin{aligned}
M_{13} &= \frac{2 \eta}{3} [R_s \cos \theta - (L_d-Lq) \omega \sin \theta ], \\
M_{23} &= \frac{2 \eta}{3} [ -R_s \sin \theta + (L_d-Lq) \omega \cos \theta], \\
M_{31} &= \eta \left( R_s \cos \theta - L_d \omega \sin \theta \right), \\
M_{32} &= -\eta \left( R_s \sin \theta + L_q \omega \cos \theta \right), \\
M_{33} &= \eta R_s + R_f + \frac{2}{3}\eta^2 \dot{L}_{AA}.
\end{aligned}  \notag
\right.
\end{equation} }
Hence from this, we get the explicit ODE form
$$
\begin{aligned}
\begin{bmatrix} 
\frac{d}{dt}i_d  \\ 
\frac{d}{dt}i_q \\ 
\frac{d}{dt}i_f 
\end{bmatrix}
&= 
-\bar A^{-1}\bar B
\begin{bmatrix} 
i_d  \\ 
i_q \\ 
i_f 
\end{bmatrix} +\bar A^{-1} \bar C,
\end{aligned}
$$
where\\
\begalis{
\bar A & :=\begmat{ 
L_d & 0  & \frac{2 \eta}{3} L_d \cos \theta \\ 
0  & L_q  & -\frac{2 \eta}{3} L_q \sin \theta \\ 
\eta L_d \cos \theta  & -\eta L_q \sin \theta  & \frac{2}{3}\eta^2 L_{AA} 
}, \\
\bar B &:=\begmat{ 
Rs & -\omega L_q  & M_{13} \\
\omega L_d  & Rs  & M_{23} \\ 
M_{31}  & M_{32} & M_{33}
},\\
\bar C&:=\begmat{ 
v_d  \\ 
v_q - \omega \phi\\ 
\eta \phi \omega \sin \theta
}.
}

\end{document}